\documentstyle[epsf,psfig]{mn}

\def\Mdot{\hbox{$\dot M$}} 
\def\Msun{\hbox{$M_\odot$}} \outer\def\gtae
{$\buildrel {\lower3pt\hbox{$>$}}   \over  {\lower2pt\hbox{$\sim$}} $}
\outer\def\ltae    {$\buildrel         {\lower3pt\hbox{$<$}}     \over
{\lower2pt\hbox{$\sim$}} $}
\def\rchi{{${\chi}_{\nu}^{2}$}}

\title[Indirect imaging of the accretion stream of V895 Cen]  
{Indirect imaging of the accretion stream in eclipsing polars IV: V895 Cen}
\author[N. Salvi et al.]  {Nikita Salvi$^1$,
Gavin Ramsay$^1$, Mark Cropper$^1$, D. A. H. Buckley$^{2}$, 
R. S. Stobie$^{2}$\\   
$^1$Mullard Space Science Laboratory, University College London,
Holmbury St. Mary, Dorking Surrey, RH5 6NT\\
$^2$South African Astronomical Observatory, PO Box 9, Observatory
7935, South Africa}

\begin{document}

\label{firstpage}

\maketitle

\begin{abstract}

We present spectroscopic and high speed photometric data of the
eclipsing polar V895 Cen. We find that the eclipsed component is
consistent with it being the accretion regions on the white dwarf.
This is in contrast to Stobie et al who concluded that the eclipsed
component was not the white dwarf. Further, we find no evidence for an
accretion disc in our data. From our Doppler tomography results, we
find that the white dwarf has $M$\gtae 0.7\Msun.  Our indirect imaging
of the accretion stream suggests that the stream is brightest close to
the white dwarf. When we observed V895 Cen in its highest accretion
state, emission is concentrated along field lines leading to the upper
pole. There is no evidence for enhanced emission at the magnetic
coupling region.

\end{abstract}

\begin{keywords}

accretion--binaries: eclipsing--stars:magnetic fields--cataclysmic 
variables--stars: individual: V895 Cen

\end{keywords}

\section{Introduction}

Polars are interacting binaries consisting of a red dwarf secondary
and a strongly magnetized ($\sim$10-200 MG) white dwarf primary. In
these systems the secondary fills its Roche lobe. Material falls under
gravity from the secondary towards the primary, initially along the
binary orbital plane before the magnetic field of the primary forces
it to leave the orbital plane and eventually impacts quasi-radially
onto the white dwarf.  Unlike non-magnetic cataclysmic variables (CVs)
the strong magnetic field of the white dwarf is high enough to prevent
the formation of an accretion disc around the primary.

Although our understanding of the accretion flow near the white dwarf
is now relatively well understood (eg Wu 2000 and references therein)
the region where accretion flow first interacts with the magnetic
field of the white dwarf is not. The flow interacts with the
magnetosphere in a complex manner and it is not easy to isolate stream
emission from other emission sources in the system (which are
generally brighter).

Eclipsing systems provide an opportunity to study the accretion flow
as a separate and distinct source for a short period of time, when the
emission from the bright accretion region on the white dwarf is
blocked by the secondary. Light curves of the eclipse contain
information about the structure and the brightness distribution along
the stream. The stream brightness distribution can be retrieved using
an indirect imaging technique which can reconstruct the brightness of
the region between the primary and the secondary.

One such technique is that of Hakala (1995) who devised an indirect
imaging method based on Maximum Entropy to deduce the brightness
distribution along the accretion stream of HU Aqr. This technique has
been developed further by Harrop-Allin et al (1999a, 1999b, 2001) who
used a more physically realistic stream trajectory, improving the
model's optimizing algorithm and including the projection effects.

V895 Cen was discovered serendipitously using the {\sl EUVE} satellite
(Craig et al 1996). Craig et al observed strong line emission and
low/high brightness states and concluded V895 Cen was likely to be a
polar. Its orbital period of $P_{orb}$=4.765 h (Stobie et al 1996) is
the second longest period currently known for a polar and the system
is found to alternate frequently between high and low accretion
states. Stobie et al (1996) concluded that it is an eclipsing polar
but that the secondary minimum of the ellipsoidal variation was offset
with respect to the eclipse.  They suggested the eclipsed component
was a hot compact source which appeared to be distinct from the white
dwarf, probably associated with the accretion stream. Howell et al
(1997) also concluded the eclipse to be of an extended object much
larger than the white dwarf, perhaps a partial accretion disc which
forms during the high state. Stobie et al (1996) found no evidence for
significant levels of polarization in a low accretion state.

To investigate the nature of the eclipsed component, we have applied
the techniques of indirect imaging and Doppler tomography to this
system to study the spatial and temporal changes in the stream and to
determine if the eclipse is associated with the white dwarf or the
stream.

\section{Spectroscopic observations}

Low resolution spectra of V895 Cen were obtained with the goal of
detecting the presence of cyclotron humps and hence the magnetic field
strength of the accreting white dwarf from their spacing. Medium
resolution spectra were also obtained with the goal of mapping the
location of the line emission regions in the binary system using
Doppler tomography.

\subsection{High accretion state observations}

Spectra were obtained of V895 Cen on 2 and 3 March 1998 using the ANU
2.3 telescope and the double beam spectrograph. The conditions were
good and the seeing was $\sim1-2^{''}$. Blue and red spectra were
taken using the 300 l/mm gratings and the effective wavelength range
were 3800--5500\AA\hspace{2mm} and 6400--8700\AA\hspace{2mm}
respectively. Exposure times were 240 sec. With a slit width of
2.0$^{''}$ the FWHM of the lines in the arc lamp spectra were measured
as $\sim$5\AA\hspace{2mm} near H$\alpha$.

On the first night V895 Cen was observed from $\phi\sim$0.94--0.30 (on
the ephemeris of Stobie et al 1996) while on the second night we were
not able to accurately attach time information due to a computer
problem. It was clear, however, that V895 was observed in a high
accretion state with prominent emission lines of H and He. There is no
convincing evidence for the presence of cyclotron humps in our
spectra.

\subsection{Low accretion state observations}
\label{low}

More spectra were obtained of V895 Cen on 18 April 1998 using the ANU 2.3
telescope and the double beam spectrograph. Some light cloud was
present at the start of the night and also towards the end of the
night. The seeing was $\sim1^{''}$.  Blue and red spectra were taken
using the 600 l/mm gratings resulting in effective wavelength ranges
of 3800--5600\AA\hspace{2mm} and 5800--7400\AA. Exposure times were
520 sec and 500 sec for the blue and red spectra respectively. With a
slit width of 1.2$^{''}$ the FWHM of the lines in the arc lamp spectra
were measured as 2.1 \AA\hspace{2mm} near H$\alpha$. A total of 51
fully calibrated spectra were obtained in each arm of the
spectrograph.

\begin{figure}
\begin{center}
\setlength{\unitlength}{1cm}
\begin{picture}(8,5.5)
\put(-2.5,-28.5){\includegraphics{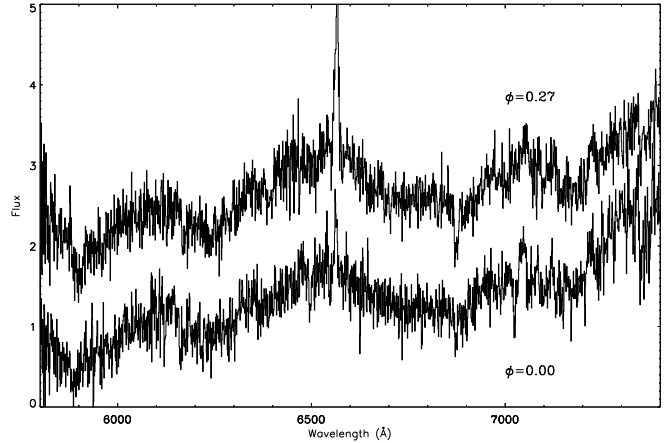}}
\end{picture}
\end{center}
\caption{Spectra taken using the ANU 2.3m double beam spectrograph on 18 April
1998 when it was in a low accretion state. Our spectra are phased on the
ephemeris in Stobie et al (1996) and we show spectra from two binary orbital
phases. The upper spectrum has been displaced vertically by 1.0 flux unit
(which is arbitrary). At both phases the spectrum is dominated by the secondary
star.}
\label{spec} 
\end{figure}

\begin{figure}
\begin{center}
\setlength{\unitlength}{1cm}
\begin{picture}(8,9)
\put(-2,7){\includegraphics{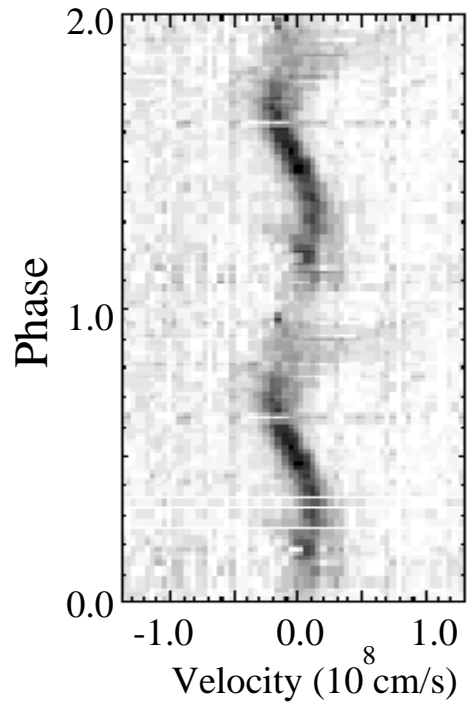}}
\end{picture}
\end{center}
\caption{The H$\alpha$ emission line taken on 18 April 2000 folded on the
binary orbital period.}
\label{fold} 
\end{figure}

V895 Cen was found to be in a low accretion state in contrast to our
March 1998 observations. The
signal to noise ratio of the blue spectra were very low and so we do
not discuss these any further. Two of the red spectra are shown in
Figure \ref{spec}. The spectrum taken during $\phi$=0.0 is almost
entirely that of the dwarf secondary star. Comparing template spectra
of late type dwarf stars we estimate the spectral type of the
secondary to be M1--M2 -- consistent with the M2 estimate of Stobie et
al (1996). Figure \ref{spec} also shows the spectrum at $\phi$=0.27.
Apart from increased emission at H$\alpha$ the shape and intensity of
this spectrum is virtually unchanged from that at $\phi$=0.0. We show
the wavelength variation in the spectrum near H$\alpha$ over the
binary cycle in Figure \ref{fold}.

\section{Doppler Tomography}
\label{tomo}

Doppler Tomography is used to map the accretion flow in binary systems
in velocity space. It has been applied to several polars such as HU
Aqr (Schwope, Mantel \& Horne 1997). Here we use the code of Spruit
(1998) to produce a Doppler map of V895 Cen. Many Doppler maps of
polars show emission from the irradiated face of the secondary
star. We can place some constraint on the mass ratio, $q=M_{2}/M_{1}$,
by varying $q$ and $M_{1}$ so that emission is located in the correct
location. We can constrain 0.25\ltae$q$\ltae0.7 purely from these
maps.

We can constrain $M_2$ using the secondary mass-orbital period
relationship (Warner 1995) and obtain $M_{2}\sim$0.46\Msun. Although
there is some evidence that secondaries in CVs with binary orbital
periods greater than 3 hrs have spectral types later than implied by
the solar-abundance main sequence for isolated stars, the spectral
type for V895 Cen found in \S \ref{low} is consistent with that
expected for its binary orbital period (Beuermann et al 1998). Based
on theoretical relationships between the radius of the secondary and
binary orbital period (Baraffe \& Kolb 2000) we estimate the
uncertainty on $M_{2}$ as $\sim$0.05\Msun for $P_{orb}$=4.765 hr. 

\begin{figure}
\begin{center}
\setlength{\unitlength}{1cm}
 \begin{picture}(8,8.5)
\put(0,-6.5){\includegraphics{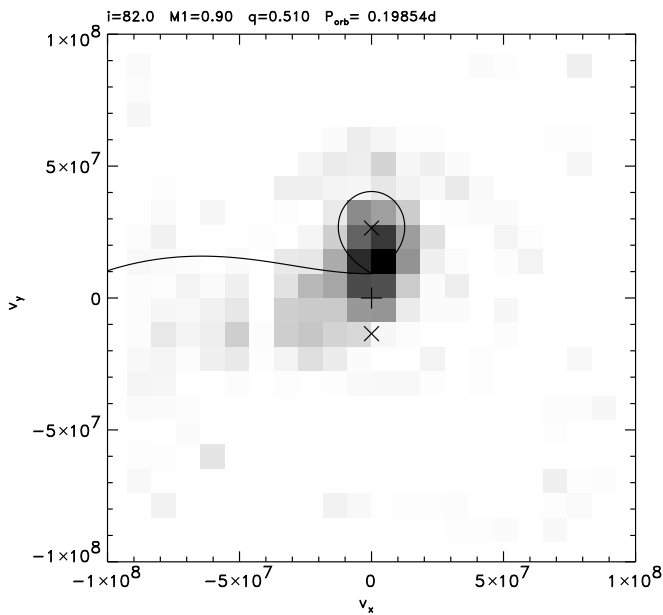}}
\end{picture}
\end{center}
\caption{The Doppler map of V895 Cen obtained in April 1998 when it was in a
low accretion state made using the H$\alpha$ emission line. The solid curve
represents the predicted path of the ballistic trajectory of the accretion
stream leaving the L$_{1}$ point. We have assumed $M_{1}$=0.9\Msun and
$M_{2}$=0.46\Msun.}
\label{dopplermap} 
\end{figure}

In generating our Doppler map we assume a binary inclination of
82$^{\circ}$ (\S \ref{eresults}). For $M_{2}$=0.41--0.51, we find that
$M_{1}>$0.7\Msun if most of the line emission is located at the
irradiated face of the secondary. We show the Doppler map of V895 Cen
in H$\alpha$ emission in Figure \ref{dopplermap} for $M_{1}$=0.9\Msun
and $M_{2}$0.46\Msun.  We use a mass of $M_1 = 0.9M_\odot$ for the
rest of the paper, implying $q$=0.51.

Although our Doppler map shows that most of the line emission in
H$\alpha$ originates at the irradiated face of the secondary star,
there is faint emission from a stream component. Compared to systems
in a high accretion state (eg HU Aqr), the stream is located at lower
$V_{y}$ than the predicted ballistic trajectory of the stream. There
is no evidence for an accretion disc in these spectroscopic data.

\section{Photometric observations}

Photometric observations of V895 Cen were made in 1997 March and May
at the Sutherland site of the South African Astronomical Observatory.
The March data were obtained using the 1.0m telescope and the May data
using the 0.75m telescope. In both cases the detector was the UCT CCD
camera in frame transfer mode (see Stobie et al. 1996). Integration
times were 4 and 10 sec for the March data and 20 sec for the May data
and the data are all filter-less (ie `white-light'). After flat
fielding and bias subtraction the images were processed using the
DoPHOT package (Mateo \& Schechter 1989).  Both differential PSF-fit,
and aperture, magnitudes were derived (instrumental), corrected for
extinction using frame standard stars.  On-chip binning (2 x 2) was
used in conditions of poor seeing. The magnitudes are instrumental and
we have used the ephemeris of Stobie et al.~(1996) to phase the data.

\subsection{General light curve features}

\begin{figure}
\begin{center}
\setlength{\unitlength}{1cm}
\begin{picture}(8,12)
\put(-2,-2){\includegraphics{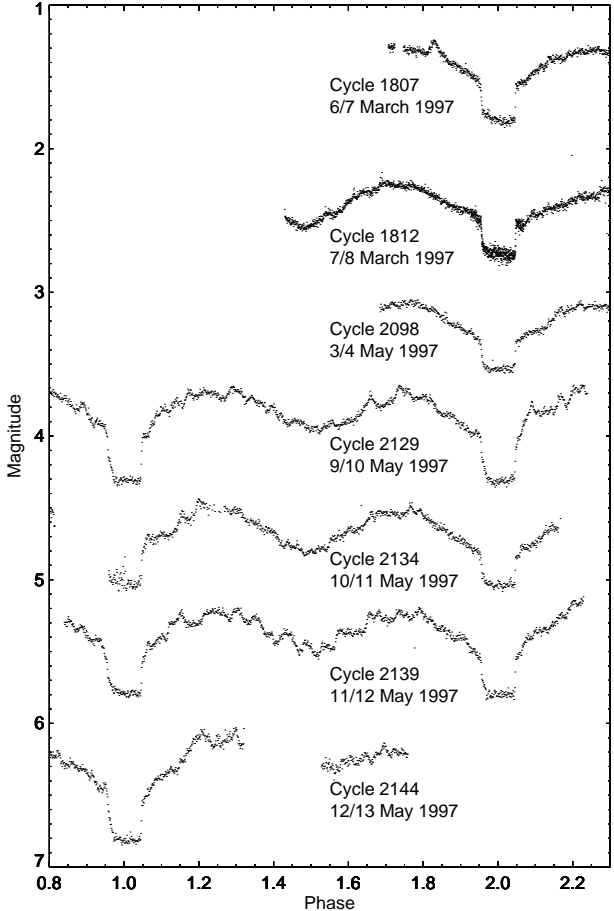}}
\end{picture}
\end{center}
\caption{Light curves obtained from the 1997 March 6 to 1997 May
12. The curves have been offset along the y-axis for clarity (since
they are instrumental magnitudes the offsets are arbitrary). The cycle
number refers to the number of cycles elapsed since the start of the
ephemeris of Stobie et al (1996). The integration times for cycle 1807
was 10 sec, 4 sec for cycle 1812 and 20 sec for the May observations.}
\label{allcycle} 
\end{figure}

During our photometric observations in March--May 1997, V895 Cen was
observed to increase from an intermediate/low to a high accretion
state (Figure \ref{allcycle}). This is evident from the increase in
the amplitude variation over the orbital cycle and the increased depth
of the eclipse. Also, in the observations made in March, there is little
evidence of flickering, while flickering becomes quite prominent in
cycles 2139 and 2144.

\subsection{The eclipse profiles}

\begin{figure}
\begin{center}
\setlength{\unitlength}{1cm}
\begin{picture}(8,16)
\put(-2,-3){\includegraphics{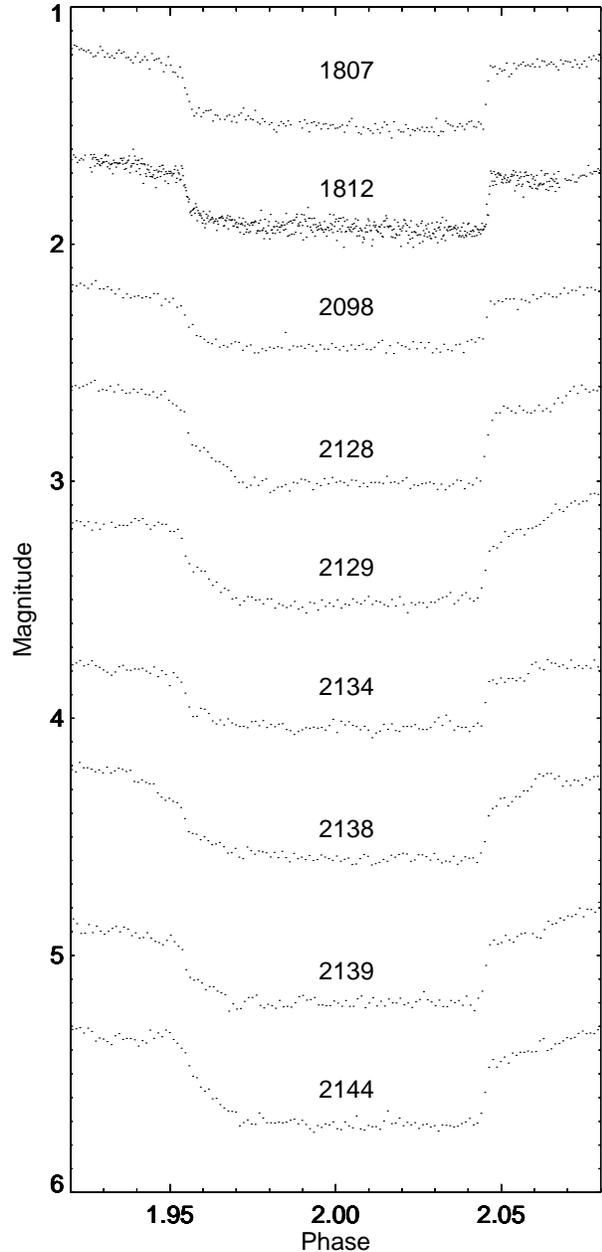}}
\end{picture}
\end{center}
\caption{As for Figure \ref{allcycle} but focusing on the eclipse profiles.
The integration time for cycle 1807 was 10 sec, cycle 1812 4 sec and 20 sec for
the other cycles.}
\label{eclipse} 
\end{figure}

In Figure \ref{eclipse} we show the light curves centered on the
eclipses. The duration, and start and end phase of the eclipse, is
constant, taking $\sim$26.6 min. This is consistent with the duration
of the low/intermediate state data ($\sim$26.3 min) of Stobie et al
(1996). In cycles 1807 and 1812, we observe a rapid eclipse ingress,
which takes $\sim$40 sec and for the eclipse egress $\sim$20 sec. In
the succeeding eclipses, the duration of the eclipse ingress becomes
progressively longer as a result of the accretion stream brightening,
though there is still a rapid decrease in intensity present at the
same phase in most cycles.

The resolution of these data is not high enough to resolve the ingress
and egress of the accretion region(s) on the white dwarf, which
typically take 1--2 sec (cf Perryman et al 2000).  We now consider if
the duration of the eclipse is consistent with the parameters derived
in \S\ref{tomo} ($q=0.5$, $i=82^{\circ}$ and $M_{1}=0.9$\Msun).  We
assume the Nauenberg (1972) mass-radius relationship for a white dwarf
and the secondary is a main sequence star. We determine the size of
the eclipsed source by tracing the Roche potential out of the binary
system along the line of sight from any point in the vicinity of the
white dwarf. We can then measure the eclipse duration for these
parameters: they are consistent with the observed eclipse duration. We
find no evidence in either these data or in the low state data of
Stobie et al (1996) for the presence of an accretion disc. The
variable duration of the ingress is due to the gradual eclipse of the
accretion stream. As the system becomes brighter, the stream remains
visible for a longer duration after the eclipse of the white dwarf.

\section{Eclipse mapping}

In the eclipse mapping model of Harrop-Allin et al (1999a, 1999b,
2001) various assumptions have to be made. These include defining the
system parameters (such as the masses of the primary and secondary
stars), the path of the accretion flow, and whether accretion is
occurring onto one or both magnetic poles. The model takes into
account the fact that the stream is optically thick using a simple
projection factor (the sine of the angle between the line of sight and
the tangent to the stream at each point).  We assume point-like
accretion regions located at the foot-points of the accreting field lines of the
white dwarf. In addition the light curves need to be corrected for emission
originating from components other than the accretion stream.  We
address these issues in turn.

\subsection{System parameters}

The parameters, inclination, $i$, the coupling radius at which the
stream becomes attached to the magnetic field, $R_{\mu}$, the angle
between the spin axis and the magnetic axis (the magnetic co-latitude)
$\beta$ and the magnetic longitude $\zeta$ of the dipole field line
are fitted in the model fit. Since it is an eclipsing system,
$i>$72$^{\circ}$ for the above system parameters. From the results in
\S 3 we fix $q$=0.51.

\subsection{The smoothing term}

Another key parameter used in the fitting process is $\lambda$, the
Lagrangian multiplier, which determines how smooth the final solution
is (Harrop-Allin 1999a).  If $\lambda$ is set too low the model fits
noise in the data. If $\lambda$ is set too high the model smoothes
small scale features and the resulting stream images have poor
resolution. The appropriate value for our data with 4 and 10 sec
resolution (cycles 1807 $\&$ 1812) was found to be $\lambda = 10^5$
and for the data with 20 sec resolution $\lambda = 10^4$. For
comparison, Harrop-Allin et al (1999a) found that $\lambda = 10^5$ was
appropriate for their test data with 0.7 sec resolution.

\begin{figure}
\begin{center}
\setlength{\unitlength}{1cm}
\begin{picture}(8,14)
\put(-2.5,-1){\includegraphics{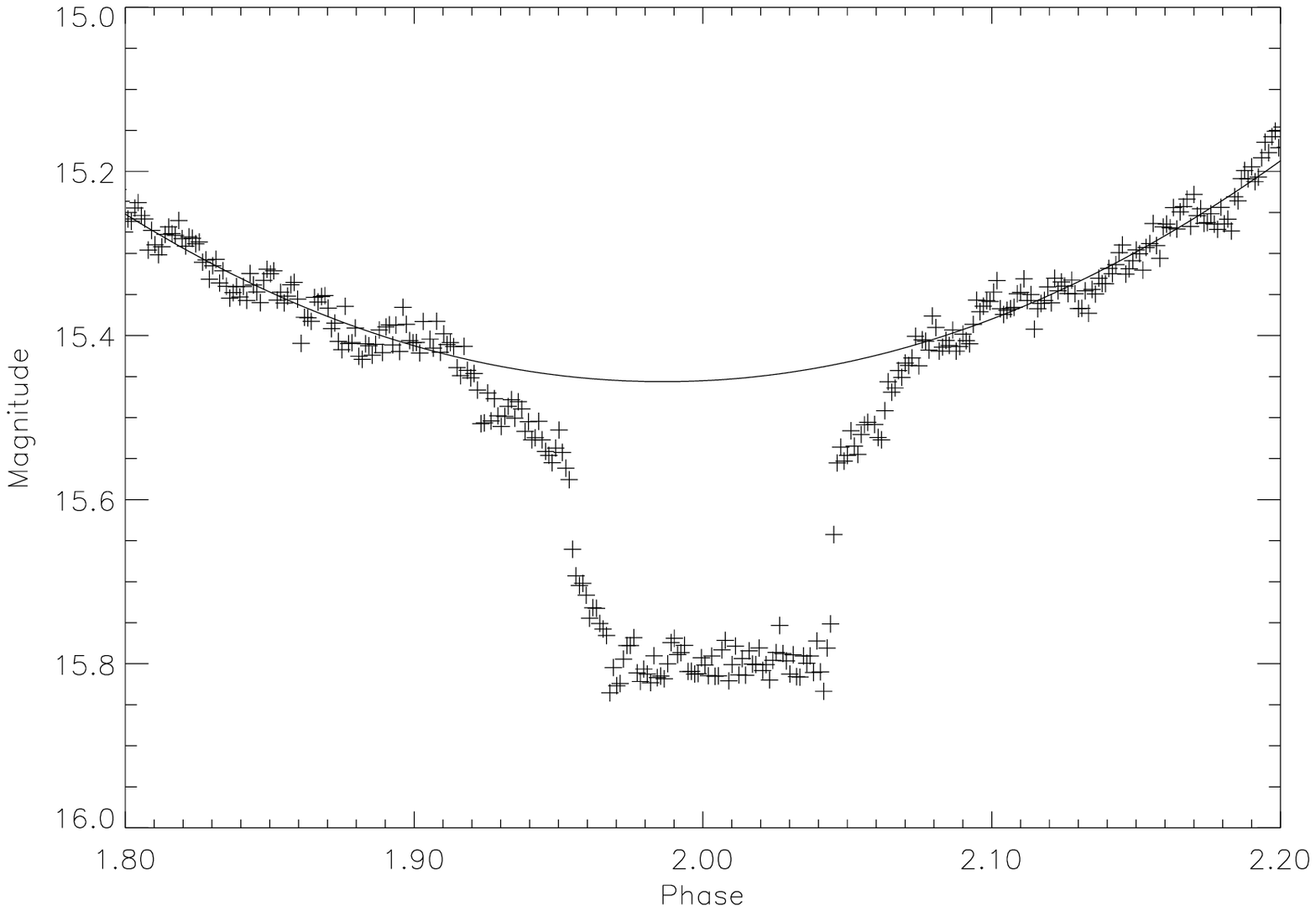}}
 \put(-2.5,-8.3){\includegraphics{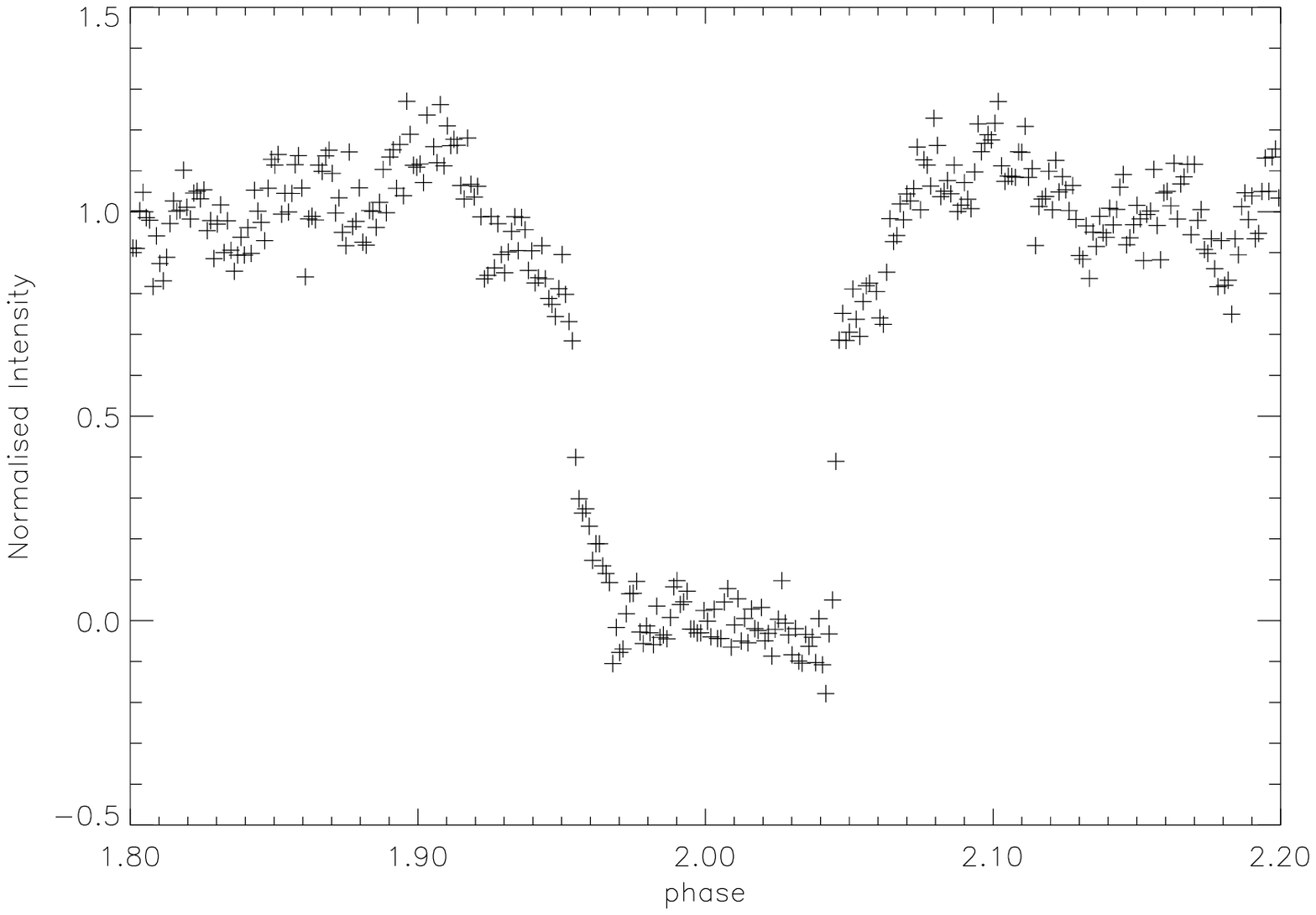}}
\end{picture}
\end{center}
\caption{Top panel shows a polynomial fit (degree=2) to the region of the
light curve near the eclipse in cycle 2139. The bottom panel shows the
normalized eclipse profile.}
\label{polyfit} 
\end{figure}

\subsection{The stream trajectory}

One of the assumptions of the model of Harrop-Allin et al (1999a) is
the path of the accretion flow: the stream path consists of a
ballistic trajectory from the $L_1$ point to the threading region,
after which the stream is forced to follow the magnetic field lines of
the white dwarf out of the orbital plane. If the stream deviates
significantly from this trajectory, then poor fits to the data are
expected.

\subsection{Removing non-stream emission}

The most important sources of flux in the optical band are the
changing viewing angle of the accretion regions on the white dwarf
(since cyclotron radiation is strongly beamed), and the accretion
stream.  The optimum method to subtract the cyclotron component would
be to have simultaneous optical polarimetry data since cyclotron
emission is polarized. In the absence of such data we have used a very
simple approach to subtract the contribution from the accretion
region. A polynomial (degree=2) was fitted to the light curves between
$\phi\sim$0.8--1.2 (data from the actual eclipse phases were
excluded). The resulting fit was then subtracted from the original
data. The normalized eclipse profiles were used as input to the
imaging method. The polynomial fit and the results of the orbital
trend removal for cycle 2139 are shown in Figure \ref{polyfit}.  To
check if the resulting stream images are sensitive to the way the
normalization was made, we also fitted a polynomial to the light curve
between other phase intervals (eg $\phi\sim$0.9--1.1). The resulting
stream images were found to be very similar.

Observations of other polars (eg UZ For, Bailey \& Cropper 1991) show
that the white dwarf photosphere is much fainter in optical light
compared with the accretion region. We therefore do not add a separate
component to account for the contribution of the white dwarf
photosphere to the overall flux.

In our modeling, large errors were assigned to data with phases
$\phi<$0.85 and $\phi>$1.15 since the accretion stream is fully
visible and no useful information can be obtained. This also applies
to data occurring at the sharp ingress and egress ($\phi$=0.952 and
$\phi$=1.046). This is due to the fact that the model assumes a point
source for the emission region rather than a region of any extent.

\subsection{One or two pole accretion}

It is also necessary to specify whether the stream accretes onto one
or both foot points of the magnetic dipole field line. Harrop-Allin et
al.~(1999a) found that if a one-pole model is applied to a two pole
geometry the resulting fit will be very poor. Applying a two-pole
model to a one-pole geometry gave fits with high $\chi^2/N$ but produced stream
images with brightness concentrated on the upper stream. Given data of
sufficient quality, it is therefore possible to determine whether to
use a one-pole geometry or a two-pole geometry simply on the basis of
the model fits alone. We performed model fits using both one pole and
two pole geometry. In the one pole geometry poor fits were obtained,
$\chi^2/N \sim 10-20$. In contrast, a two pole geometry gave good
fits, with $\chi^2/N \sim 1$. A two-pole geometry was therefore
applied.

\section{Eclipse mapping results}
\label{eresults}

The goodness of the model fit to the data is more strongly dependent
on some parameters than others. The duration of the eclipse is
controlled mainly by $q$ and $i$. For various combinations of ($q,
i$), equally good fits were obtained. We found that $q$=0.51 and
$i=82^\circ$ (cf \S 3) gave good fits to data from all the cycles: we
therefore fixed these parameters at these values. For these
parameters, the angle of the magnetic axis, $\beta$, was found to be
$\beta=11^{\circ}\pm1^{\circ}$. The magneto-spheric radius, $R_\mu$,
was primarily determined by the phase at which the stream was fully
eclipsed. This uncertainty in this parameter was estimated to be
$\pm$0.05a. The uncertainty in $\beta$ and $R_\mu$ is based on the
range of values for which $\chi^2$ remains unchanged. We caution that
similar $\chi^2/N$ values could be obtained for a larger range of
values of $\beta$ and $R_\mu$ for a different combination of input
parameters.

Changing the magnetic longitude, $\zeta$, by $\pm10^\circ$ had no
significant change on the phase at which the model reached totality. A
constant value of $\zeta=85^\circ$ has been used for imaging all
cycles. We show the best fit results of our modeling in table
\ref{tresults}: good fits were obtained to the data from all the
cycles except cycle 2139 where \rchi=1.50.

\begin{figure*}
\begin{center}
\setlength{\unitlength}{1cm}
\begin{picture}(13,21.5)
\put(-4,3.2){\includegraphics{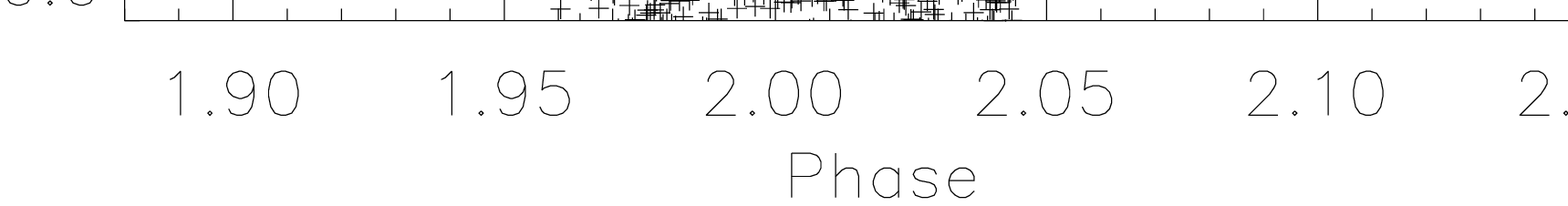}}
\put(5.5,2.2){\includegraphics{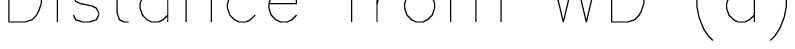}}
\put(-4,-4.3){\includegraphics{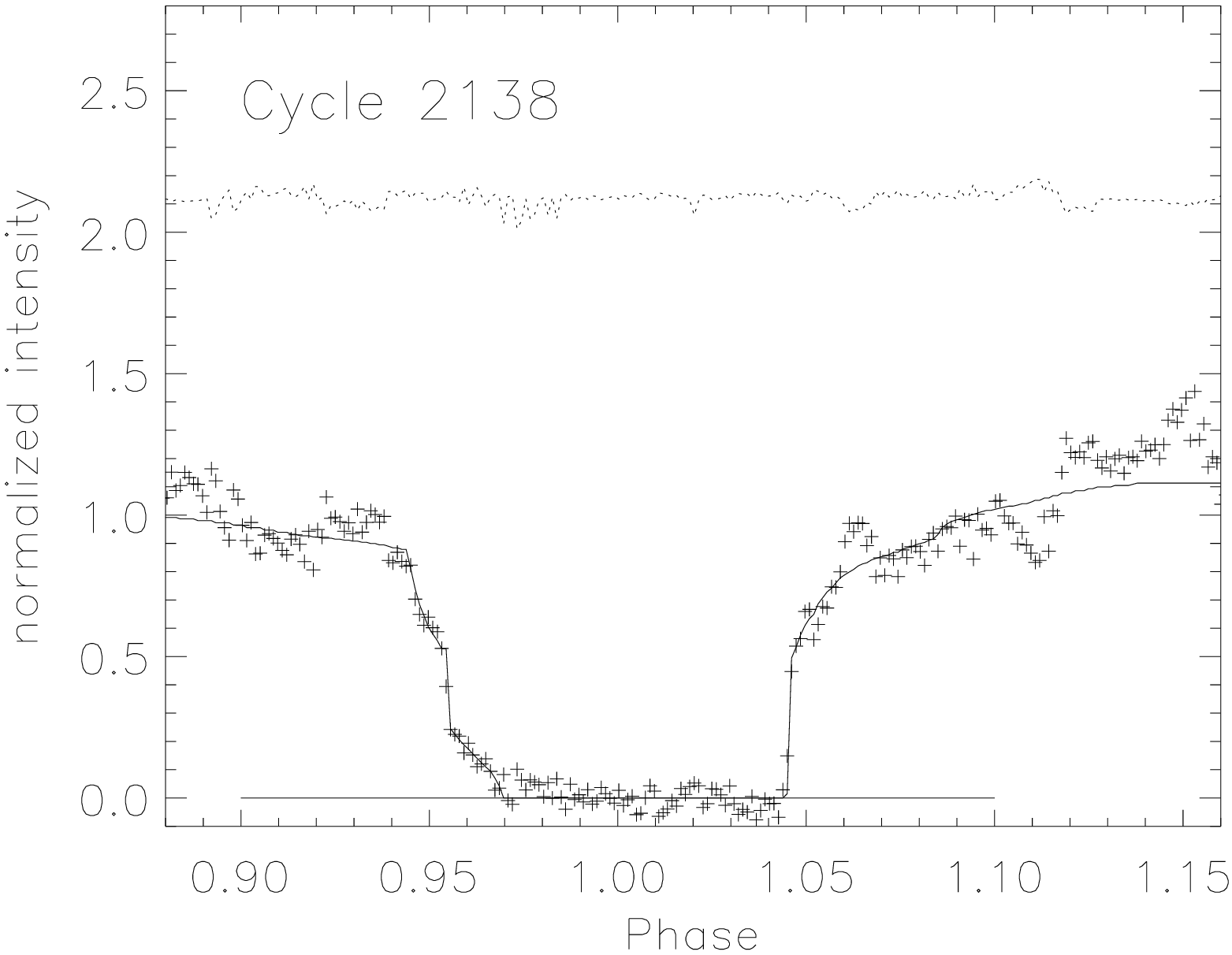}}
\put(5.5,-5.3){\includegraphics{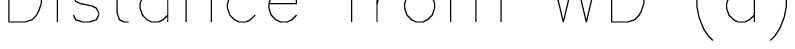}}
\put(-4,-11.8){\includegraphics{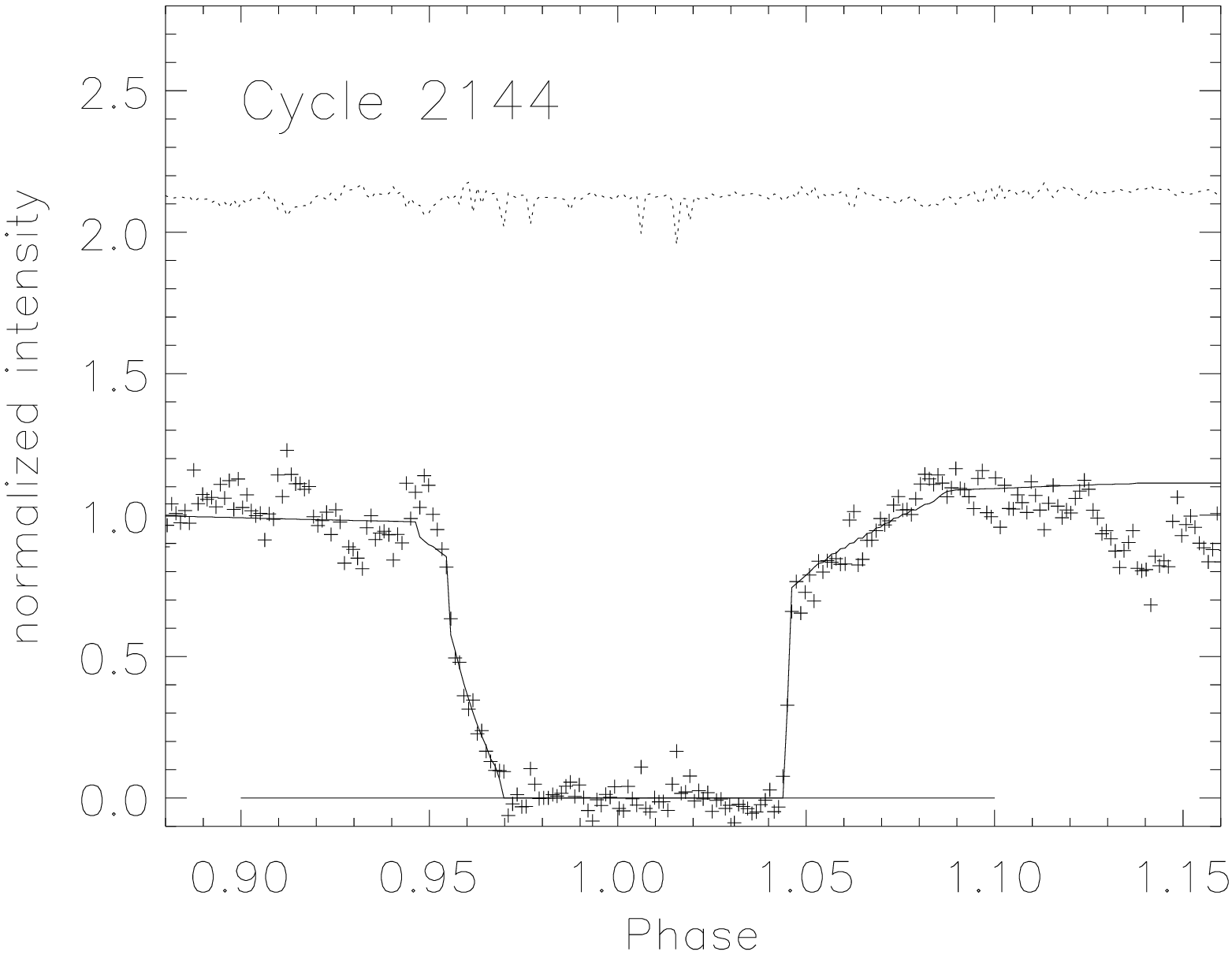}}
\put(5.5,-12.8){\includegraphics{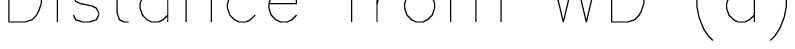}}
\end{picture}
\end{center}
\caption{Model eclipse profiles and the corresponding accretion stream
images for three orbital cycles. The stream images are shown projected
onto a plane passing through the center of both stars and
perpendicular to the orbital plane. The brightness of each emission
point is shown as a line perpendicular to the stream and passing
through the point; the brightness of the point is represented by
the length of the line. The white dwarf is shown to scale as a circle
(labeled `WD'), and the secondary is shown (not to scale) to mark the
position of the L1 point (labeled 'L1').}
\label{imaging1} 
\end{figure*}

We obtained corresponding images of the accretion stream for all our
model fits. As these stream images are similar for certain cycles we
only show three of them (along with their model eclipse profiles) in
Figure \ref{imaging1}. In selecting the images to present, we
considered which of the cycles had the best quality data and sampled
various intensity states. For instance in cycle 1812, the system was
in an intermediate/low accretion state and also had data with the
highest time resolution (4 sec) of any of our data. We also show cycle
2144 since V895 Cen was in its highest accretion state that we
observed photometrically. Intermediate in epoch and accretion state are cycles
2128, 2129, 2138 and 2139: since cycle 2138 gave the best fit (table 1), we
also show the stream map for this cycle.

In cycles 1807, 1812 and 2098, V895 Cen was in an intermediate/low
accretion state. Our results show that the accretion stream brightens
rapidly nearer to the white dwarf. (Since our photometry was
differential rather than absolute, our maps do not allow us to compare
the brightness of the stream from cycle to cycle). There is no
appreciable enhancement in emission either at $R_{\mu}$ or in the
ballistic stream component. As V895 Cen enters a higher accretion
state, the stream brightens along the full extent of the magnetically
controlled stream in both the upper and lower field lines. In our last
cycle (2144) the system is in the highest accretion state - as
inferred from the largest eclipse depth and peak to peak variation. In
this cycle, our model suggests that emission is now concentrated along
the field lines originating from the upper pole.

\begin{table}
\label{imgresults}
\begin{tabular}{@{}lccccccccc}   \hline
  & Cycle& &$\beta$ &&$\zeta$ &&$R_{\mu}$&$\chi^2/N$      \\  \hline
  &  1807 && 11.5 && 85 && 0.24a  (42 $R_{wd}$)& 0.42       \\
  &  1812 && 11.5 && 85 && 0.23a  (40 $R_{wd}$) & 0.90      \\
  &  2098 && 11.5&& 85&& 0.23a  (40 $R_{wd}$)    & 0.71     \\
  &   2128 && 10.5& &85& &0.27a  (47 $R_{wd}$)    &0.39    \\
  &   2129&& 10.0&& 85 &&0.27a  (47 $R_{wd}$)    & 0.69    \\
   &  2134 &&11.5  &&85  && 0.25a  (43 $R_{wd}$)  &0.67     \\
  &   2138 &&12.0 &&85 &&0.27a  (47 $R_{wd}$)    & 0.24     \\
  &   2139 &&12.0 &&85 & &0.26a  (45 $R_{wd}$)   & 1.50    \\
   &  2144 &&10.0  &&85  &&0.27a  (47 $R_{wd}$)  & 0.27     \\   \hline
\end{tabular}
\caption{The results of our model fits to the data. We show the best
fits to the latitude of the magnetic axis $\beta$, the magnetic
longitude, $\zeta$, and the radius from the white dwarf that the stream
gets controlled by the magnetic field, $R_\mu$. We also show the
goodness of fit. In the above results we have used $q=0.51$,
$i=82^\circ$.}
\label{tresults}
\end{table}

\section{Discussion}

\subsection{The eclipsed source}

Stobie et al (1996) suggested that the eclipsed component was not the
white dwarf. This was based on the fact that the eclipse occurred
before the secondary minimum in the light curve. Assuming the
secondary minimum was the true marker of inferior conjunction, and
taking the timings of the eclipse ingress and egress relative to this,
they concluded that the eclipsed source was $\sim$30 white dwarf radii
from the white dwarf. We have found that the phasing of the eclipse
ingress and egress occurs at exactly the same orbital phase in both
low, intermediate and high state data (cf Figure 5). If the eclipsed
source was the stream as suggested by Stobie et al (1996) we would not
expect to observe the stability in the eclipse features as we do.

We now address the fact that the eclipse appears off-set from the
secondary minimum in the low state light curves of Stobie et al
(1996). Many polars show evidence for heating of the trailing face of
the secondary by the accretion region on the white dwarf. It is
expected that even if the irradiation is sharply reduced or switched
off, the trailing face of the secondary will remain heated for some
duration. Szkody et al (1999) estimate that in the case of the polar
AR UMa it takes around 5 months for the secondary to cool down to the
temperature of the unheated part of the star. Indeed, from our Doppler
tomography results (\S \ref{tomo}) we find that the secondary in V895
Cen is still heated when we observed it in a low accretion state. This
has the effect of increasing the optical flux between $\phi$=0.5--0.9
compared to a secondary star with no irradiation.  We suggest that the
apparent offset between the eclipse and the secondary minimum is due
to asymmetric irradiation of the secondary star.

\subsection{The indirect stream mapping results}

Our results show that in the intermediate/low accretion state the
stream brightens rapidly as it nears the white dwarf. In our brightest
state data, we find that stream emission is concentrated mainly along
the field lines leading to the upper pole. At face value this suggests
that as the system reaches a high enough mass transfer rate, the
accretion mode goes from a two-pole to a one-pole model. It is not
clear why this would be the case, but it suggests that with an
increase in the mass transfer rate, the upper pole is now the more
favorable pole to accrete.

In all our model fits, there is no evidence for a brightening of the
accretion stream at the magneto-spheric interaction region. These
results are similar to that of the low accretion state data of HU Aqr
(Harrop-Allin et al.~ 2001). However, there was some
indication that the stream brightened at the interaction region in the
$U$ and $B$ bands. Using emission line data of HU Aqr in a high
accretion state and a different technique to that used here, Vrielmann
\& Schwope (2001) derived stream brightness maps and found a
brightening of the stream in the magneto-spheric interaction region of
HU Aqr. It is possible that the accretion state (and hence amount of
irradiation) plays an important role in determining whether the stream
is found to brighten at the interaction region.

It is interesting to compare the coupling radius that we derive from
our model fits with that of HU Aqr (Harrop-Allin 1999b, 2000). They
find that in a high accretion state, $R_{\mu}\sim$0.18a. We find a
mean value of $R_{\mu}\sim$0.25 from our fits. HU Aqr has a magnetic
field strength of 36 MG (Schwope, Thomas \& Beuermann 1993). Since we
expect the accretion flow to interact with the magnetic field when the
magnetic pressure equals the ram pressure of the flow, we predict that
the magnetic field strength of V895 Cen is significantly larger than
that of HU Aqr, other things (such as \Mdot) being equal.

\section{Summary}

We find from our photometric data that the eclipsed component is
consistent with the accretion regions on the white dwarf. Our indirect
imaging results suggest that the stream brightens close to the
magnetic poles in the low/intermediate accretion state. In the highest
accretion state we observe, emission is extended along the field lines
leading to the upper magnetic pole.

\section{Acknowledgments}

We would like to thank Kate Harrop-Allin and Henk Spruit for the use
of their software and the Australian National University for the
generous allocation of observing time.

\end{document}